\documentclass[a4paper,11pt]{article}
\pdfoutput=1 

\usepackage{jinstpub} 

\title{New veto hodoscope ANTI-0 for the NA62 experiment at CERN}

\author[a]{H.~Danielsson}
\author[b]{O.~Gavrishchuk}
\author[a]{P.~A.~Giudici}
\author[c]{E.~Goudzovski}
\author[d,1]{S.~Kholodenko,\note{Corresponding author.}}
\author[d]{M.~Kholodenko}
\author[e]{I.~Mannelli}
\author[d]{V.~Obraztsov}
\author[d]{V.~Sugonyaev}
\author[f]{R.~Wanke}

\affiliation[a]{CERN, European Organization for Nuclear Research,
\\CH-1211, Geneva 23, Switzerland}

\affiliation[b]{Joint Institute for Nuclear Research,
\\141980, 6 Joliot-Curie St, Dubna, Russia}

\affiliation[c]{The University of Birmingham,
\\Edgbaston, Birmingham, B15 2TT, United Kingdom}

\affiliation[d]{NRC ``Kurchatov Institute~-~IHEP``
\\ 142281, 1 Nauki sq, Protvino, Russia}

\affiliation[e]{Scuola Normale Superiore e INFN, Sezione di Pisa,
\\ Piazza dei Cavalieri, 7 - 56126 Pisa, Italy}

\affiliation[f]{Institut fur Physik and PRISMA Cluster of excellence, Universitat Mainz, \\ Staudingerweg 7, 55128, Mainz, Germany}

\emailAdd{sergey.kholodenko@cern.ch}

\abstract{The NA62 experiment is a fixed-target experiment at the CERN SPS. The main goal of the experiment is to measure the branching ratio of the ultra-rare kaon decay $K^{+}\to\pi^{+}\nu\bar\nu$. The NA62 detector allows also to study other rare kaon decays and to search for very weakly coupled particles of MeV-GeV mass-scale. The new ANTI-0 hodoscope is proposed and designed to veto events with charged halo particles entering the decay volume. It is now being assembled at CERN. The detector design, the performance simulation and the results of measurements with cosmic rays and test beams for the individual elements are presented. The commissioning and the first run of data-taking is scheduled for after the end of LS2 long shutdown (April 2021).}


\proceeding{INSTR20: Instrumentation for Colliding Beam Physics\\
  24~---~28 February 2020\\
  Novosibirsk, Russia}

\begin{document}
\maketitle
\flushbottom
\section{Introduction}\label{sec:intro}
\begin{figure}[h]
\centering
\includegraphics[width=\linewidth]{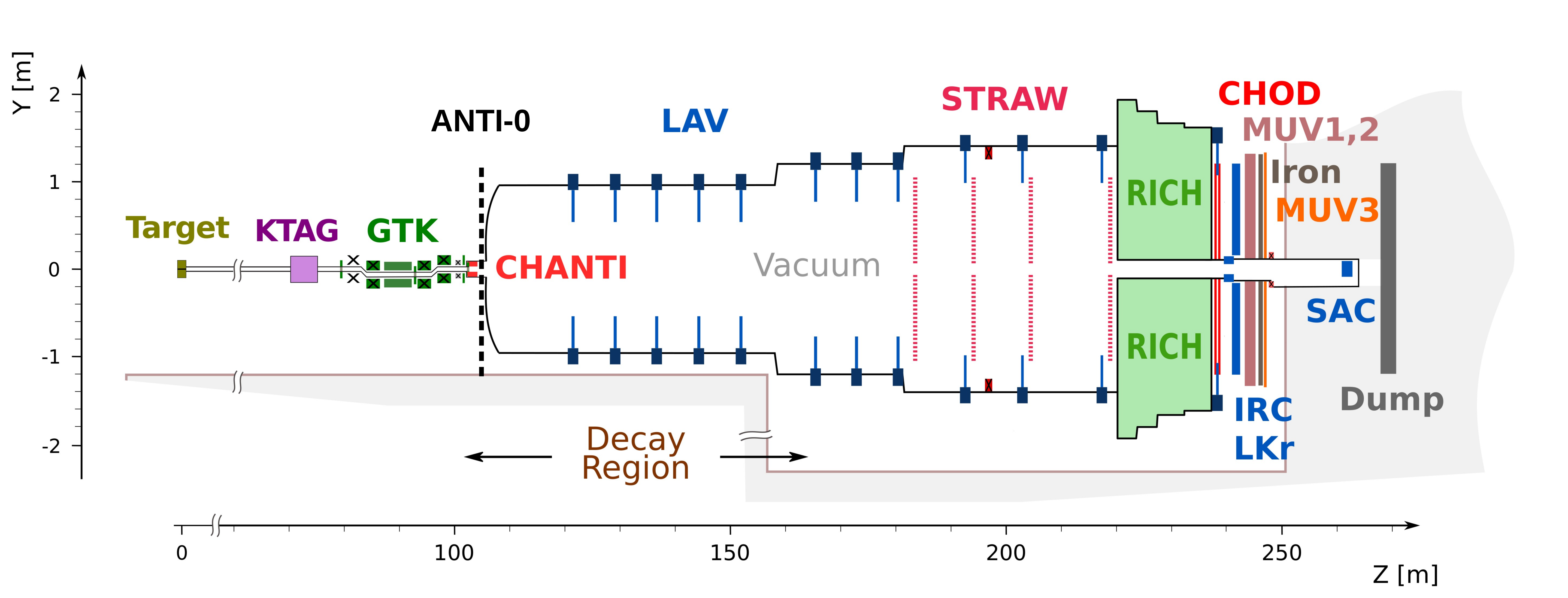}
\caption{Schematic view of the NA62 detector}\label{fig:setup}
\end{figure}

The NA62 experiment~\cite{NA62:2017rwk} is a fixed target experiment at CERN SPS. The purpose of the experiment is the study of rare and ultra-rare kaon decays, in particular the measurement of the branching ratio of the $K^{+}\to\pi^{+}\nu\bar\nu$ decay with 10\% accuracy. The 400~GeV/c proton beam from SPS is used to produce 75~GeV/c kaons. The total rate of beam particles is 750~MHz of which 6\% are kaons. The schematic view of the experimental setup is shown in figure~\ref{fig:setup}. The new veto hodoscope ANTI-0 is located in front of the Decay Volume. It is indicated with the black dashed line and has the same diameter as the vacuum tube. 
\par Kaons in the beam, concentrated in a transverse area of $30\times15$~mm$^2$, are  identified by the differential Cherenkov counter KTAG and their four-momenta are measured by the GTK spectrometer. Halo particles~(mostly muons) enter the Decay Volume at a rate of several MHz. In case they  are close in-time with a beam kaon they might be mistakenly accepted as originating from a kaon decay. The ANTI-0 should be efficient in detecting halo particles and to avoid random vetoes should have a time resolution as good as possible, definitely better than 1~ns.

\section{The ANTI-0 detector}\label{sec:Anti0}
\par
\begin{figure}[h]
\centering
\includegraphics[width=\linewidth]{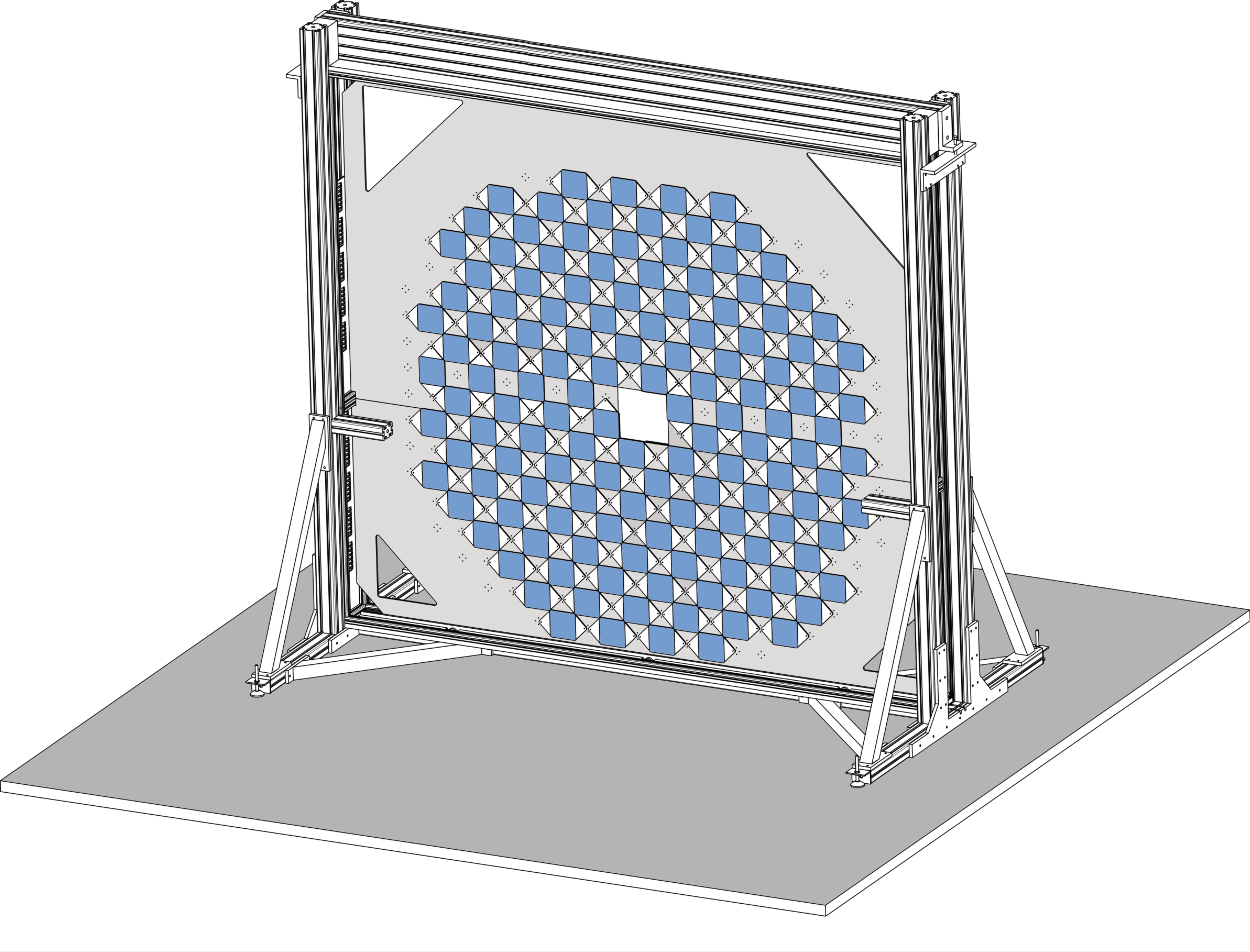}
\caption{Schematic view of the ANTI-0 detector mainframe}\label{fig:Anti0Design}
\end{figure}

 In order to achieve the specified requirements the choice for the ANTI-0 is a cell structure hodoscope covering the area $R=1080$~mm around the beam pipe and placed just in front of the Decay Volume. The detector is assembled of 280 individual counters~(figure~\ref{fig:Anti0Design}). Each counter consists of a plastic scintillator tile~(polystyrene doped with 2\% PTP and 0.02\% POPOP produced with injection-molding technique at IHEP~(Protvino)~\cite{Kadykov:1991ih}) with a size of $124 \times 124 \times 10$~mm$^3$ and is read by four S14160-6050HS SiPMs. The SiPMs are organized in two groups of two viewing the tile from two opposite $124\times10$~mm$^2$ edges. The SiPMs in one group are connected in series (for AC). The signals from the two groups of SiPMs are amplified with a gain $\sim 30$ and connected in analog ``OR``. Thus, each tile is read by a single electronic channel. Counters are wrapped with Tyvek.
\par
\begin{figure}[h]
\begin{minipage}{0.45\textwidth}
\centering
\includegraphics[width=0.95\linewidth]{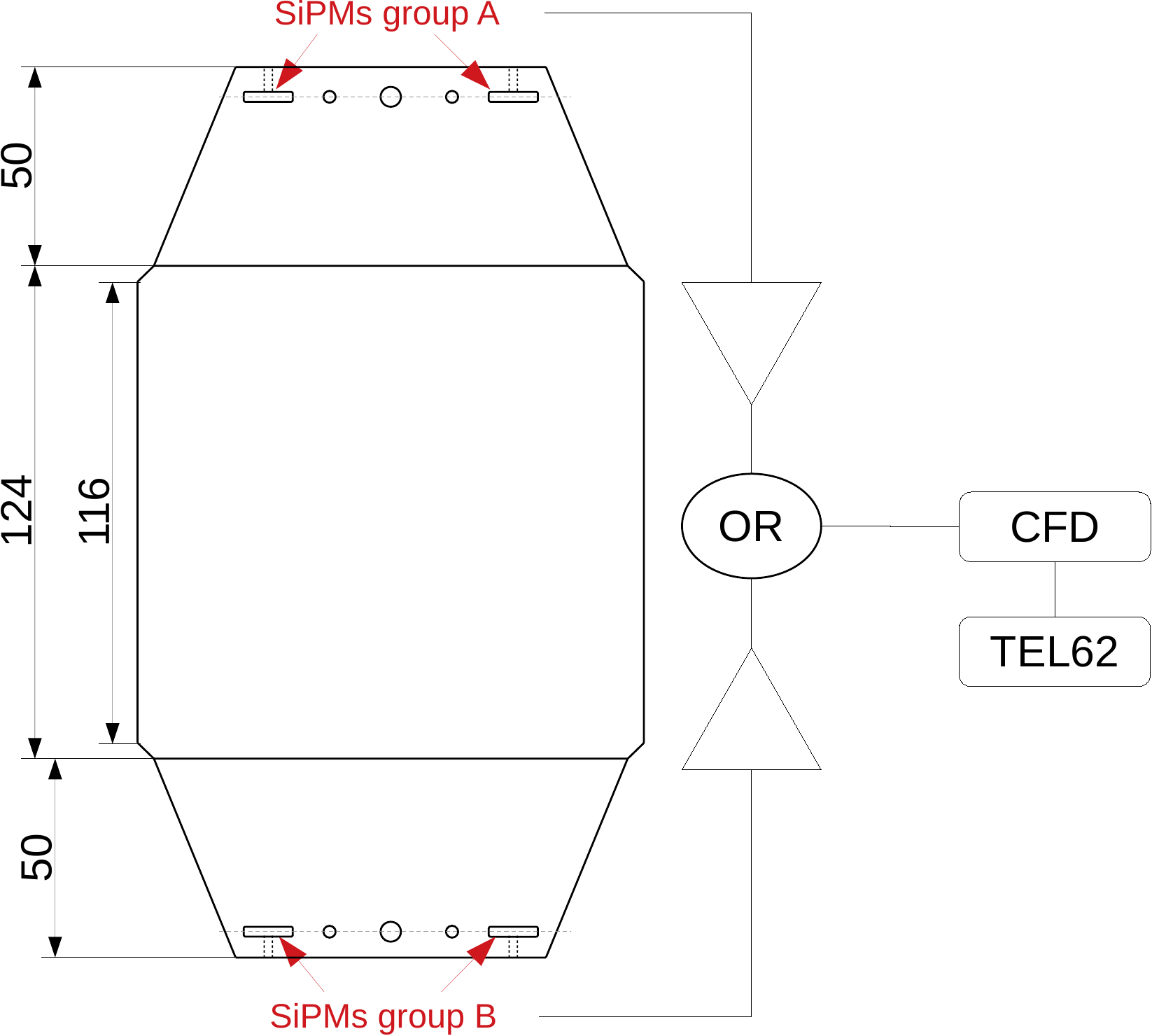}
\end{minipage}
\begin{minipage}{0.45\textwidth}
\centering
\includegraphics[width=0.95\linewidth]{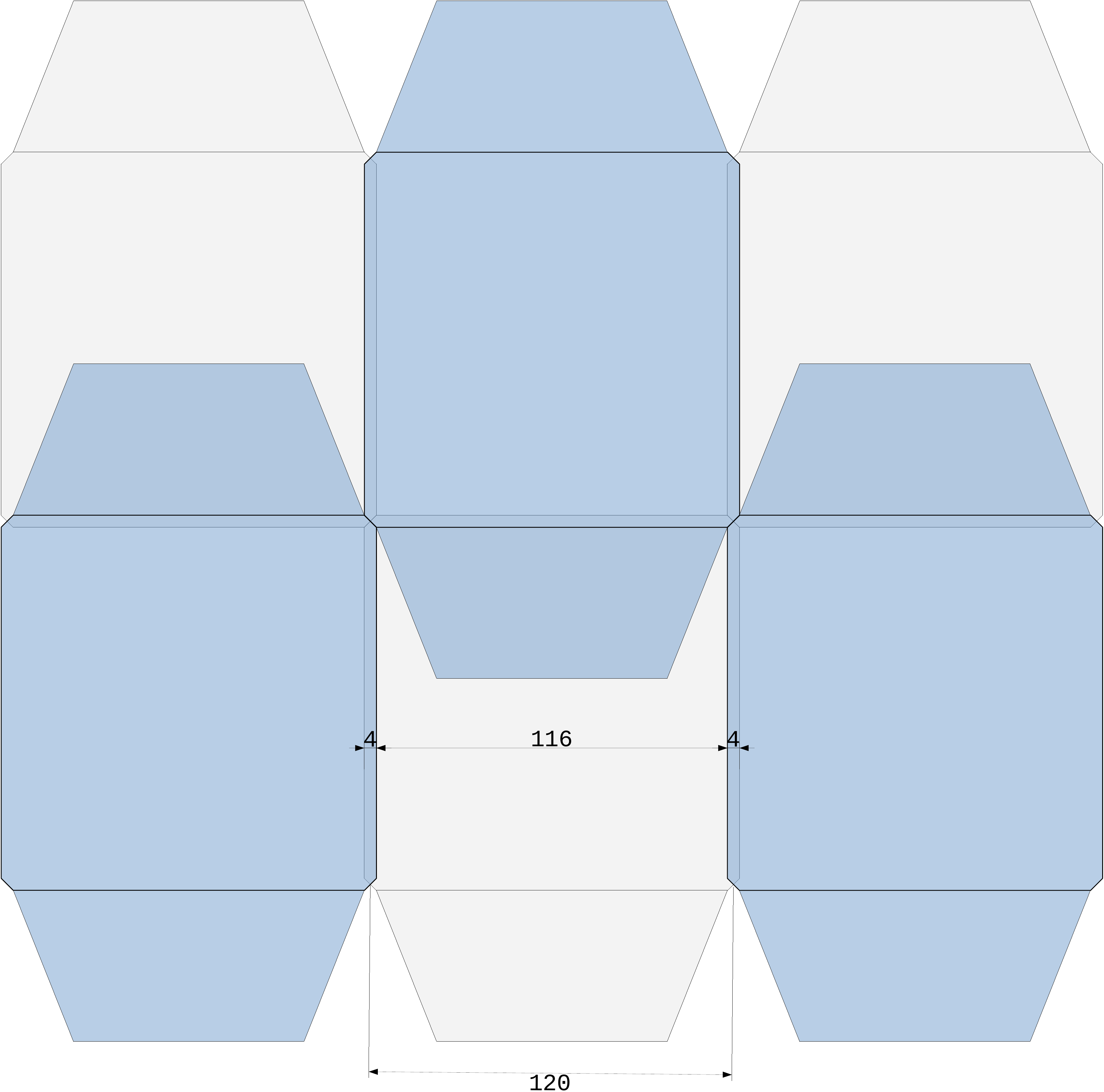}
\label{fig:Anti0Chessboard}
\end{minipage}
\caption{The schematic view of the ANTI-0 counter. Scintillating tile with two 50~mm long lightguides glued on the opposite sides~(left) and schematic view of the tiles on the mainframe~(right). Counters marked with blue and grey colours are fixed on the front and back sides of the support Al sheet. }\label{fig:Anti0Tile}
\end{figure}

 For the directly coupled SiPMs the amplitude and time of signal arrival strongly depends on the position of a charged particle crossing the counter~\cite{Kholodenko:2014kea}. To ensure the uniformity in both amplitude and timing properties the SiPMs are displaced from the edge of scintillating tile by 40~mm by using Plexiglas lightguides~(refer to section~\ref{sec:MC}). A schematic view of the basic counter is presented in the figure~\ref{fig:Anti0Tile}~(left). The four SiPMs are placed in the $2.2 \times 10$~mm$^2$ slots. Optical grease is used to ensure optical contact between the lightguide and the SiPMs which are fixed in position with nylon screws. 
\vfil
\newpage 

\par The counters are fixed on a central, 5~mm thick, Al support sheet, in chessboard style on both front and back sides~(figure~\ref{fig:Anti0Tile}~right) with a step of 120~mm that makes 4~mm overlap with all the four neighboring counters. The output signals from the counters are discriminated with constant fraction discriminators~(CFD) and recorded with TDC TEL62 modules~\cite{Ammendola:2019skv}, the NA62 DAQ boards.

\par
\begin{figure}[h]
\centering
\includegraphics[width=0.9\linewidth]{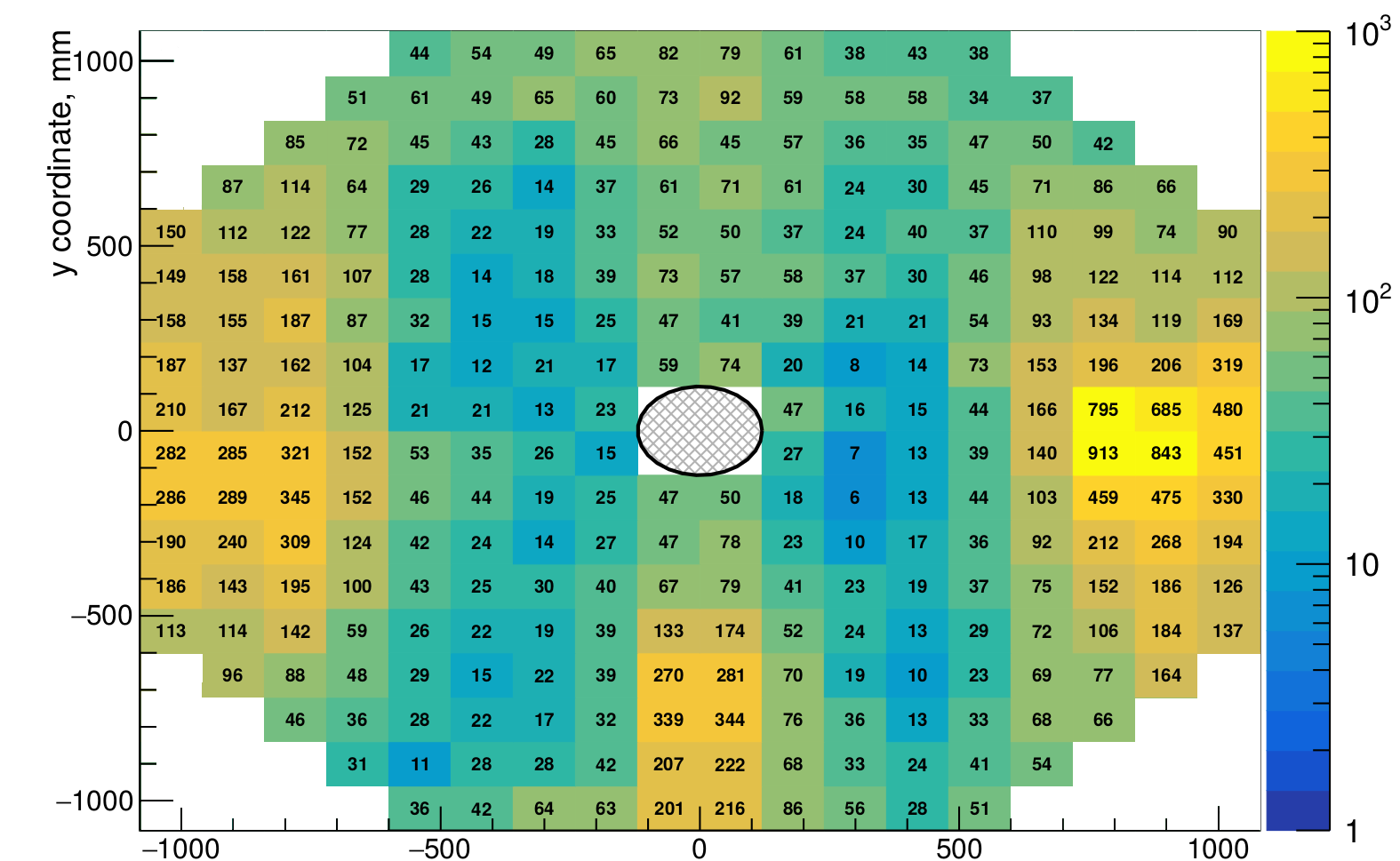}
\caption{Expected rate in kHz per tile for the ANTI-0 detector with counter size of $120 \times 120$~mm$^2$.}\label{fig:RatePerTile}
\end{figure}

 The chosen size of the counters limit the individual counting rate to 1~MHz. The expected rate~(in KHz) per counter is presented in the figure~\ref{fig:RatePerTile}.

\section{MC simulation}\label{sec:MC}
The response of individual counters to minimum ionizing particle was studied  with Geant4~\cite{Allison:2016lfl}. Optical photons, produced by Cherenkov radiation and by scintillation were propagated from the point of emission to the SiPMs positions. The shape of the pulse at the output of the OR of the preamplifiers was simulated as a convolution of the optical photon arrival time and the single photo-electron signal shape which was randomly selected from data recorded with a LeCroy WaveRunner 606Zi oscilloscope. The attenuation length of the scintillator was set as 200~cm and the SiPM photon detection efficiency~(PDE) was defined using the datasheet~(Vbias=+2.5V)~\cite{SensL}.

\subsection*{Lightguide length}
The usage of lightguides, which allow to displace photodetectors from the edge of scintillator, improves both the time and amplitude response uniformity.
\par
\begin{figure}[h]
\begin{minipage}{0.42\textwidth}
\centering
\includegraphics[width=\linewidth]{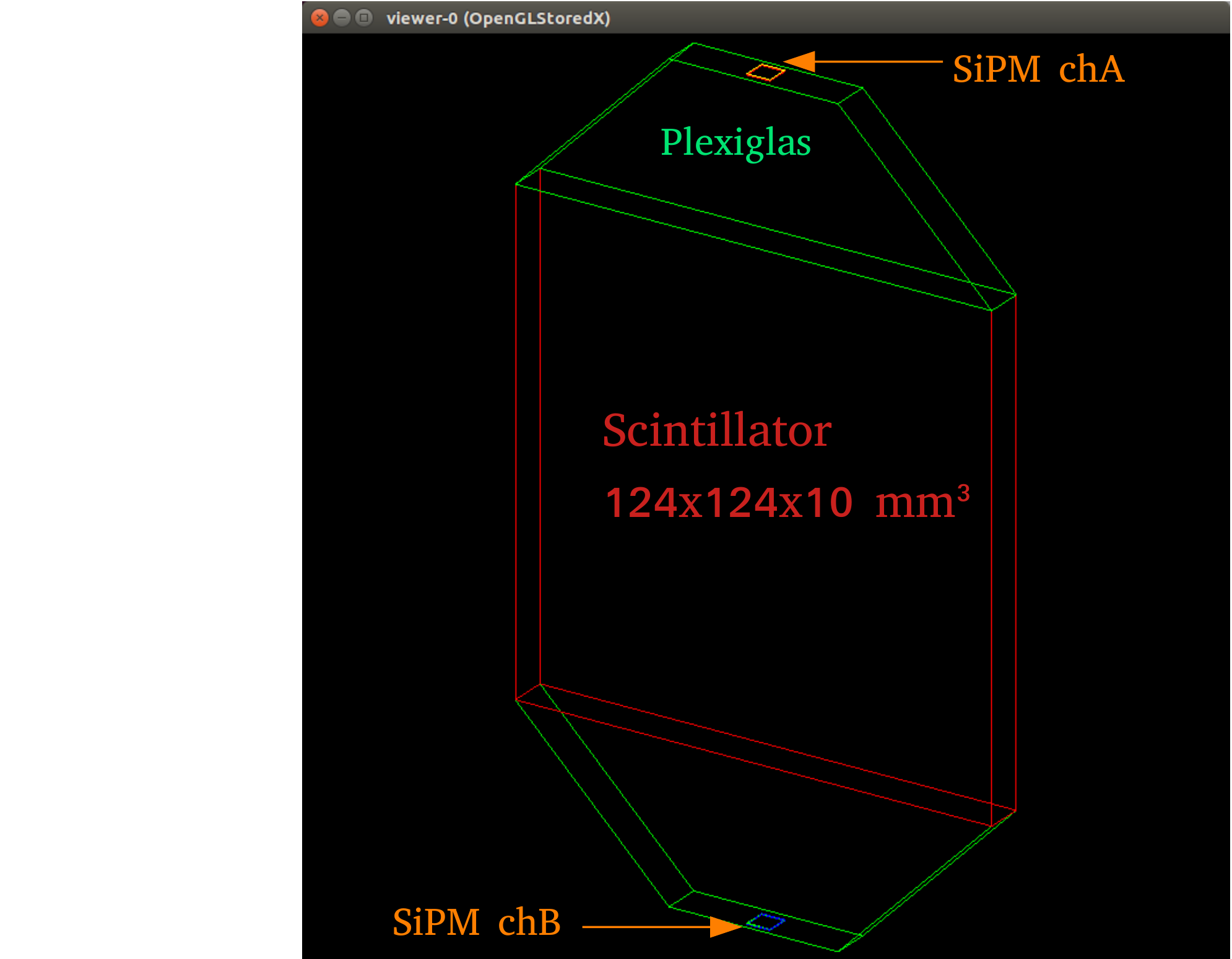}
\caption{Geant4 simplified model for the light-guide length studies.}
\label{fig:MC_LightguideLength}
\end{minipage}
\hfill
\begin{minipage}{0.5\textwidth}
\centering
\includegraphics[width=\linewidth]{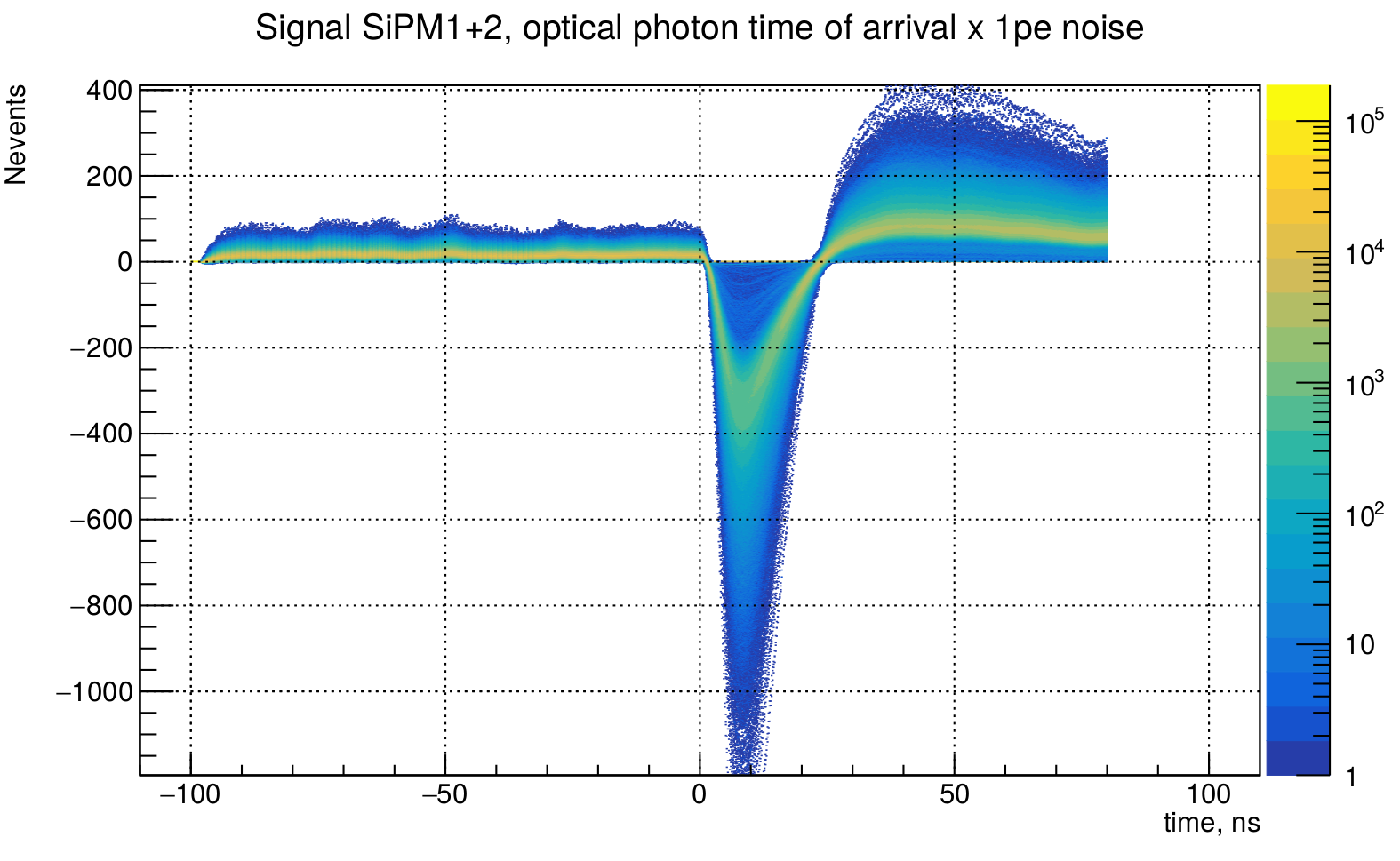}
\caption{Simulated counter response pulses~(convolution of single photon time of arrival and single photo-electron noise pulse shape).}
\label{fig:MC_TileResponce}
\end{minipage}
\end{figure}

 A simplified counter geometry~(see figure~\ref{fig:MC_LightguideLength}) was used to study the minimal acceptable light-guide length. The test counter consisted of a scintillation tile with a size $124\times124\times10$~mm$^3$ and two trapezoid Plexiglas lightguides attached to the opposite edges of the tile. The two $6\times6$~mm$^2$ SiPMs were directly coupled to the lightguides.

\par We simulated charged particles crossing the test prototype in 2~mm steps in the transverse coordinates $X$ and $Y$. For the timing studies the simulated tile pulse got discriminated using a simulated constant-fraction discriminator~(CFD) by triggering on 20\% of the peak amplitude. A time value of "0" corresponds to the charged particle crossing time. 
\par
\begin{figure}[!htb]
\begin{minipage}{0.5\textwidth}
\includegraphics[width=0.9\linewidth]{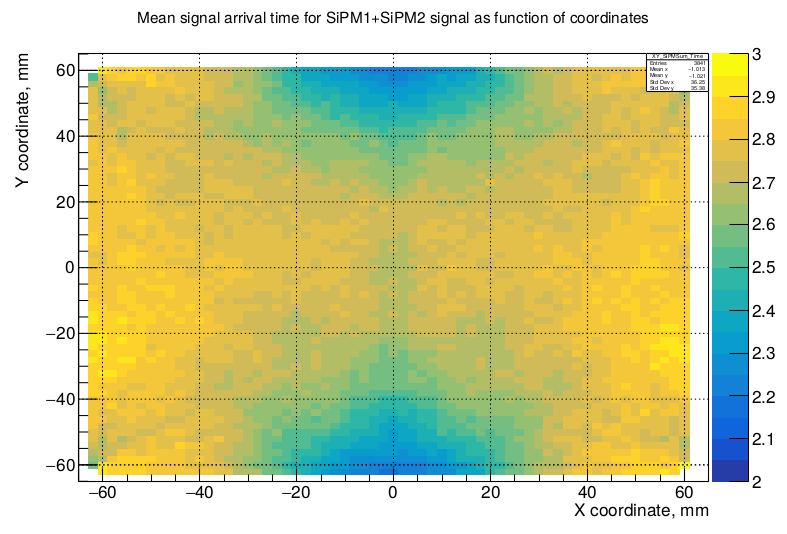}
\end{minipage}
\begin{minipage}{0.5\textwidth}
\includegraphics[width=0.9\linewidth]{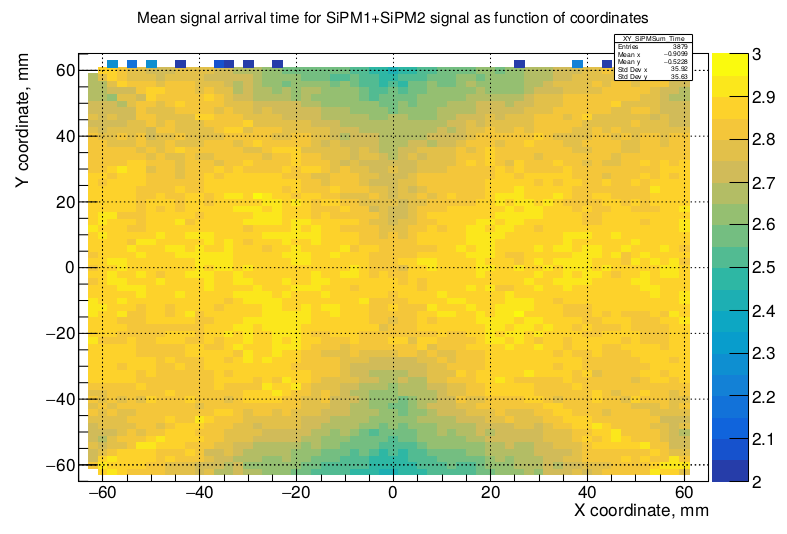}
\end{minipage}
\caption{Average time of signal arrival as a function of particle crossing coordinates for 20~mm~(left) and 40~mm~(right) lightguide length.}
\label{fig:MC_PlexLength}
\end{figure}

\begin{figure}[!h]
\begin{minipage}{0.5\textwidth}
\includegraphics[width=0.9\linewidth]{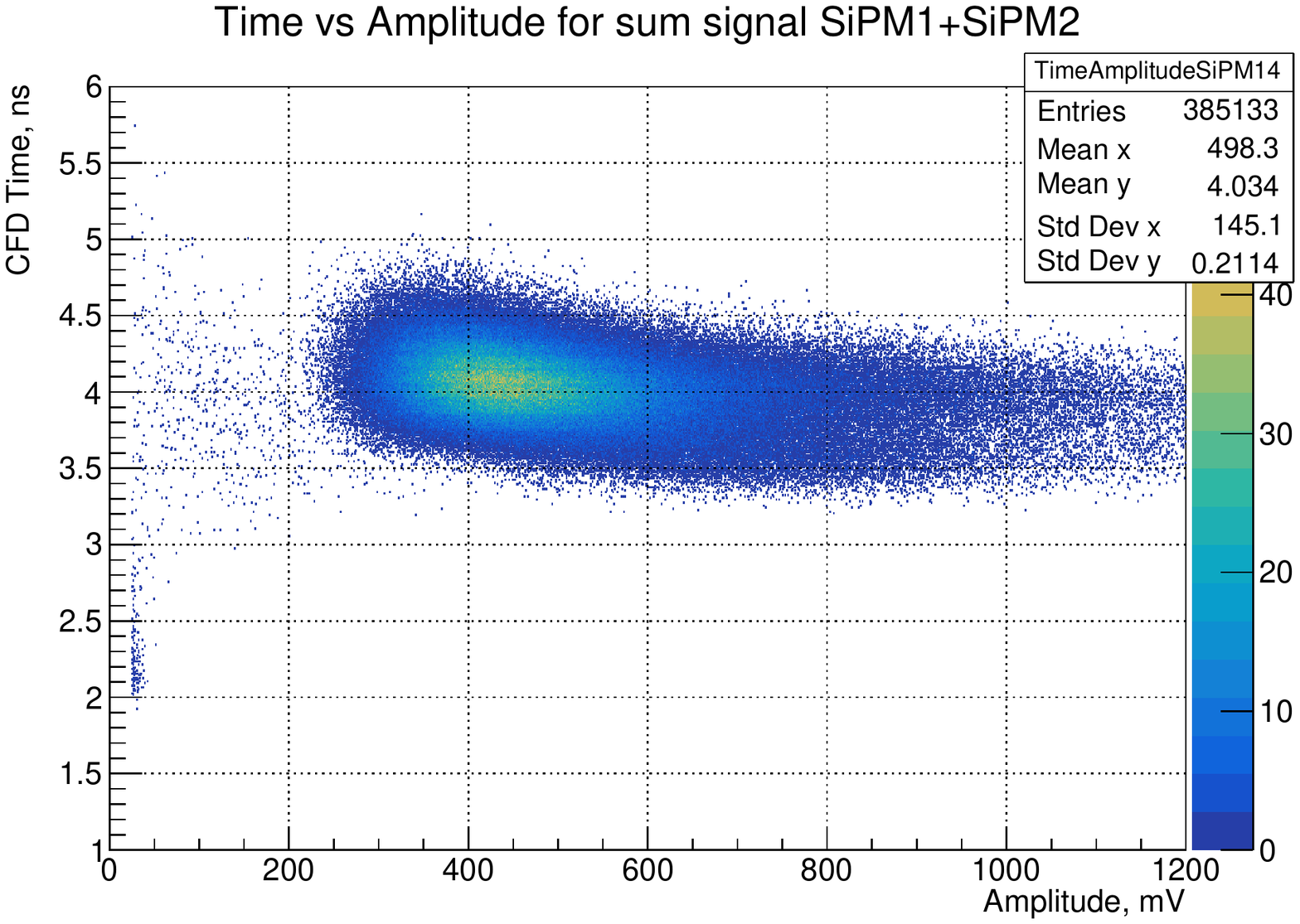}
\end{minipage}
\hfill
\begin{minipage}{0.5\textwidth}
\includegraphics[width=0.9\linewidth]{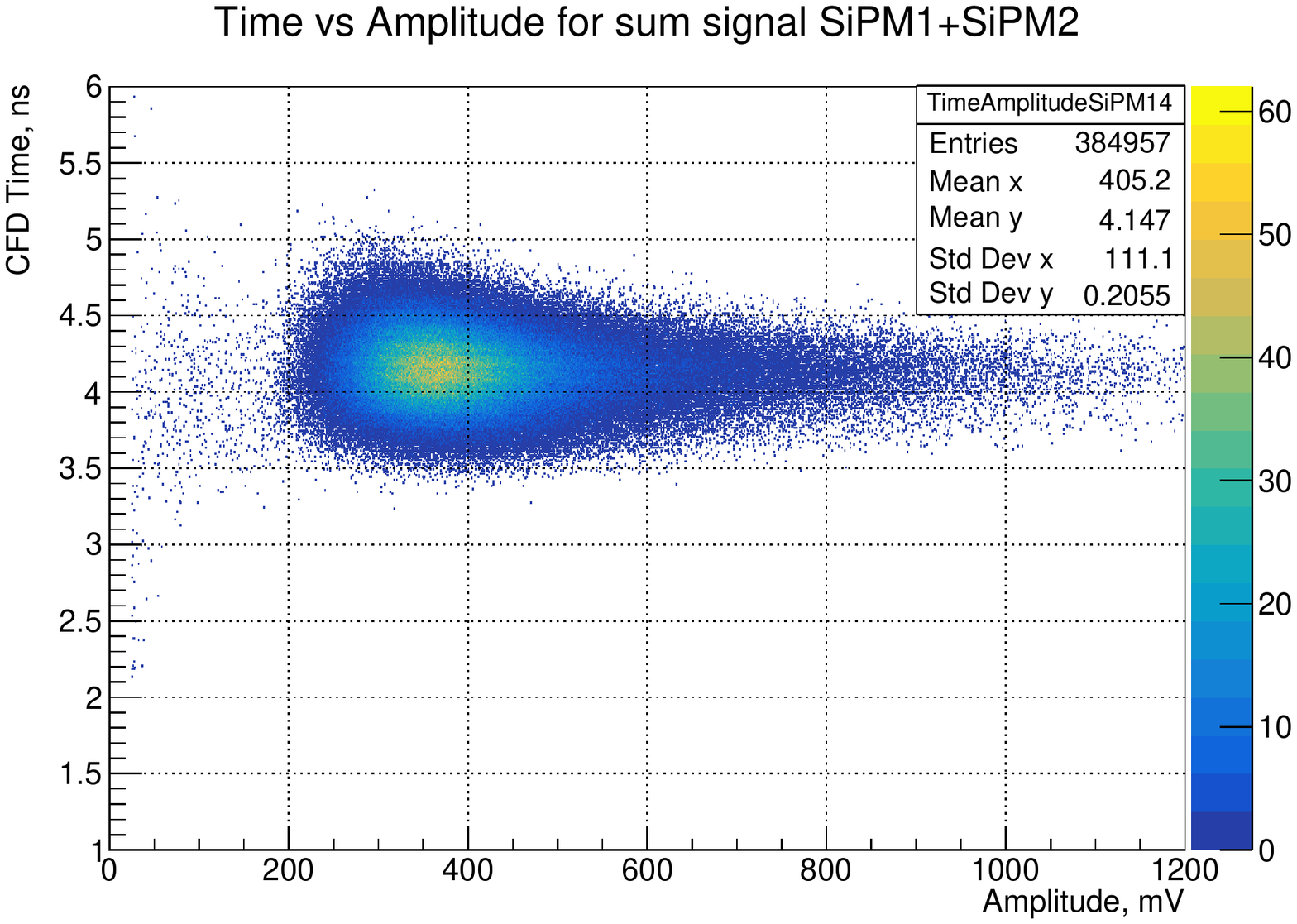}
\end{minipage}
\caption{MC simulated time of signal arrival as a function of amplitude for 20~mm~(left) and 40~mm~(right) lightguide length.}
\label{fig:MC_TAPlexLength}
\end{figure}

 Two different lengths of lightguides were considered~--- 20~mm and 40~mm. Figure~\ref{fig:MC_PlexLength} presents the average time of signal arrival as a function of particle crossing coordinates and figure~\ref{fig:MC_TAPlexLength} shows the average time of signal arrival as a function of average amplitude for options with 20~mm~(left) and 40~mm~(right) long light-guides. Notice that the maximum length of the lightguides is limited by the distance between successive tiles in the same row.
\newpage


\subsection*{Number of SiPMs per counter}
\par To further improve the uniformity and stability of the detector~(in case of one SiPM failure during the data-taking) the simulation with one, two and four SiPMs per tile is performed with fixed 40~mm lightguide length. SiPMs attached to same lightguide are placed with 60~mm distance ensuring light collection uniformity.
\par
\begin{figure}[h]
\begin{minipage}{0.32\textwidth}
\includegraphics[width=\linewidth]{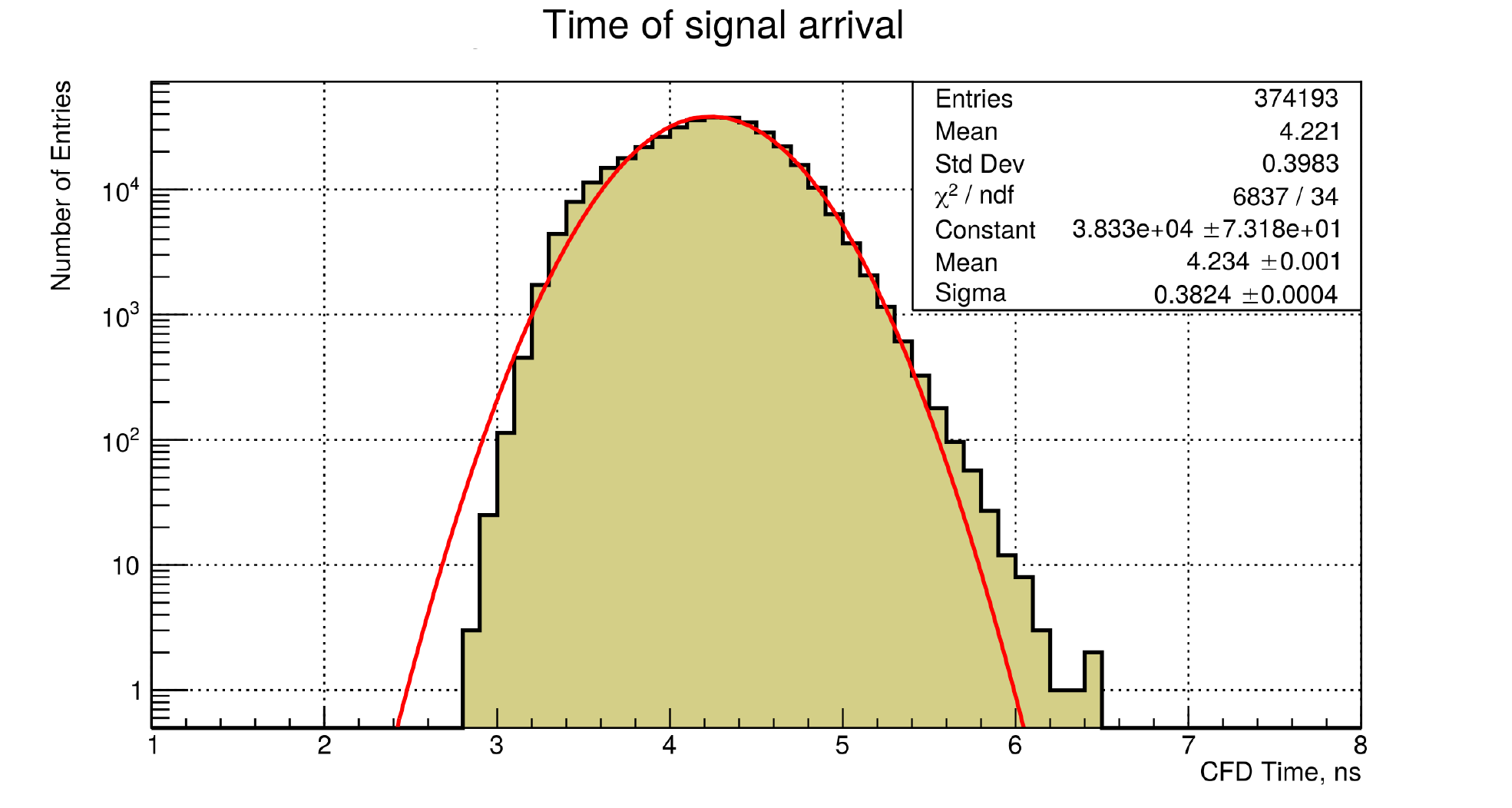}
\end{minipage}
\hfill
\begin{minipage}{0.32\textwidth}
\includegraphics[width=\linewidth]{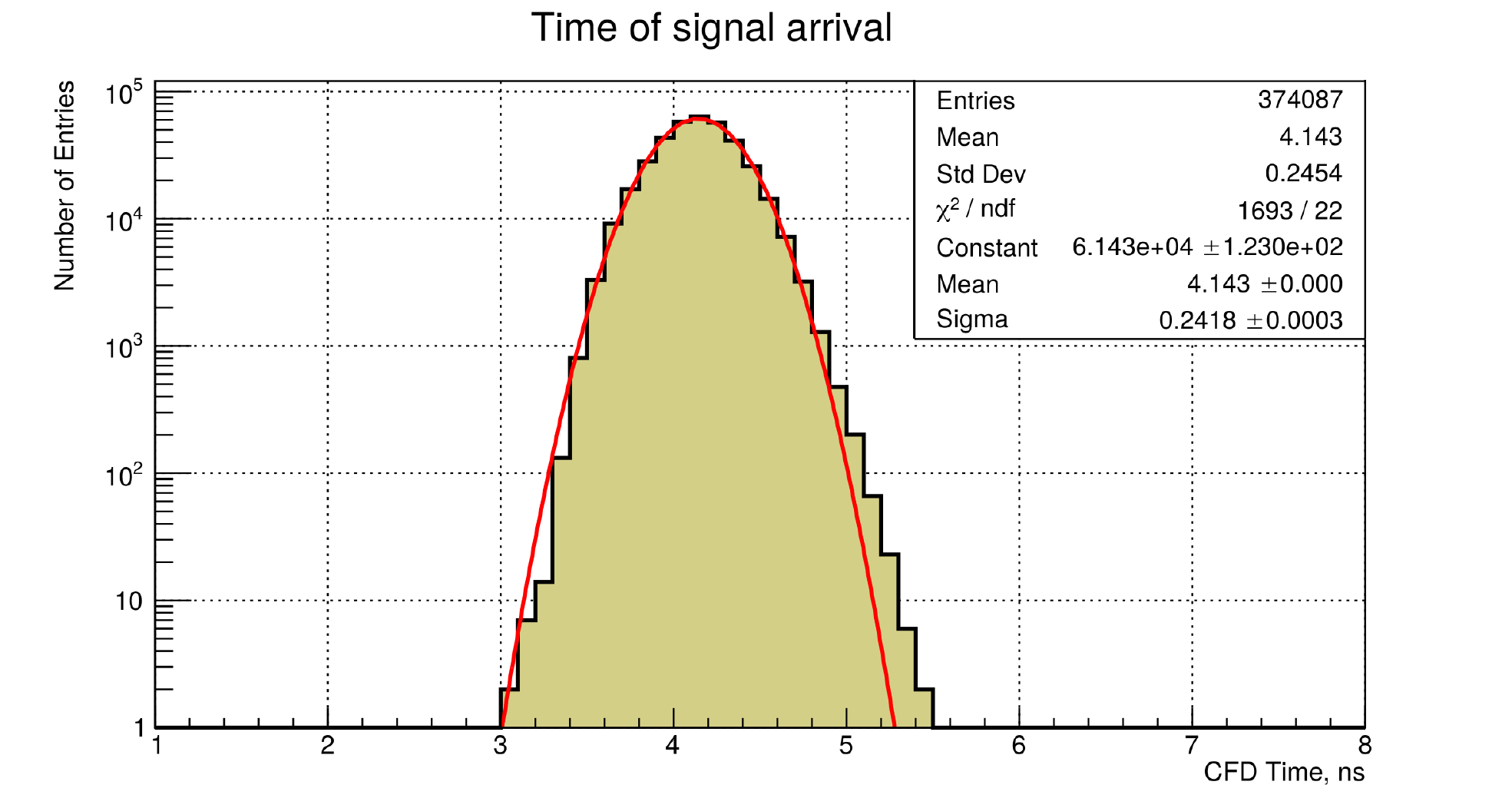}
\end{minipage}
\hfill
\begin{minipage}{0.32\textwidth}
\includegraphics[width=\linewidth]{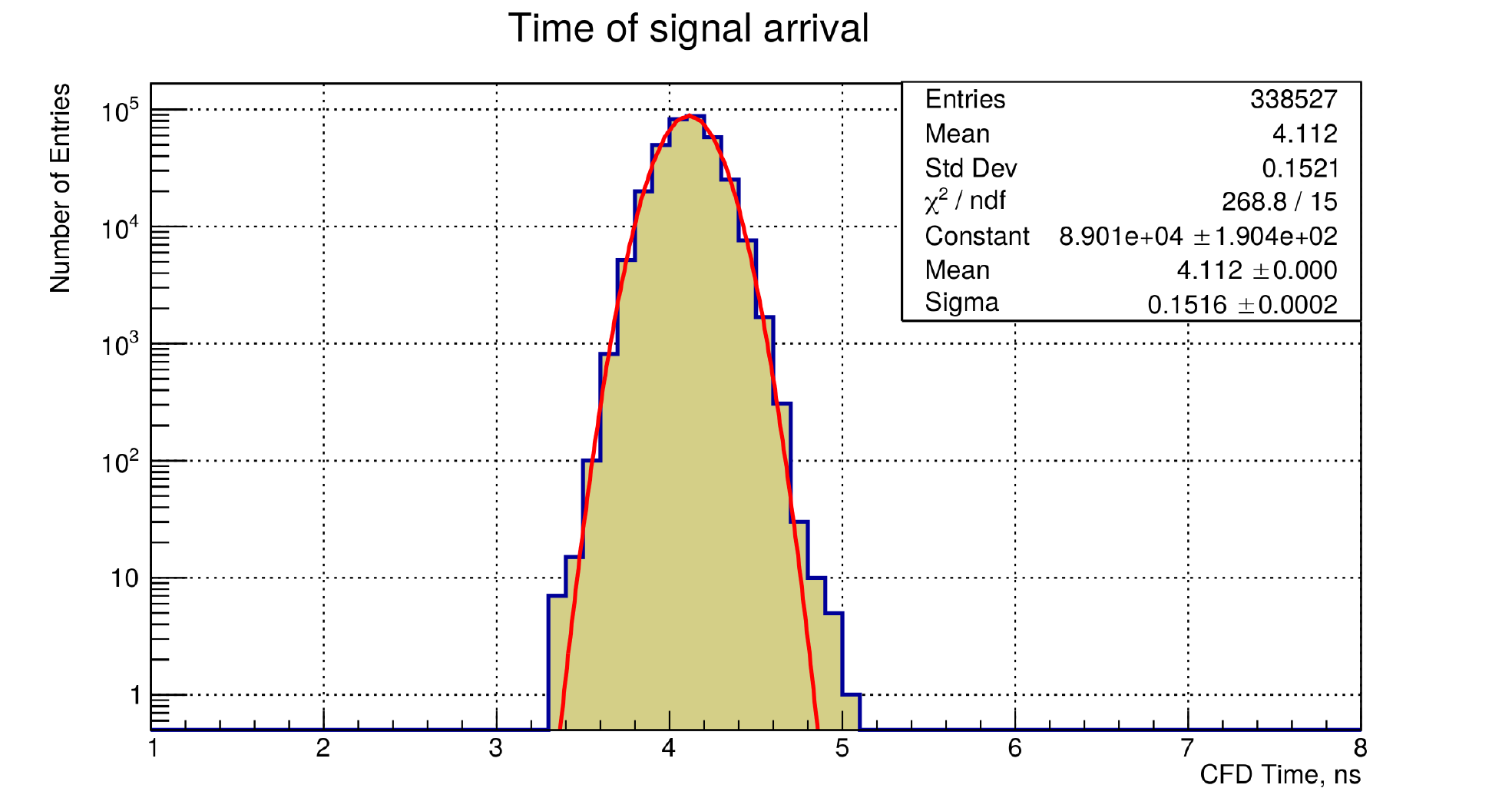}
\end{minipage}
\caption{MC simulated time of signal arrival for one~(left), two~(middle) and four~(right) active SiPMs. Events uniformly distributed over the counter surface. CFD with threshold = 0.2.}
\label{fig:MCTimeResolution}
\end{figure}

 Figures~\ref{fig:MCTimeResolution} presents distributions of time of signal arrival for three options: one~(left), two~(middle) and four~(right) active SiPMs described by Gaussian~(red line). Time resolution~($\sigma)$ improves from $380\pm5$~ns~(one active SiPM) to $150\pm5$~ps for the option with four active SiPMs.

\newpage
\par
\begin{figure}[h]
\begin{minipage}{0.32\textwidth}
\includegraphics[width=\linewidth]{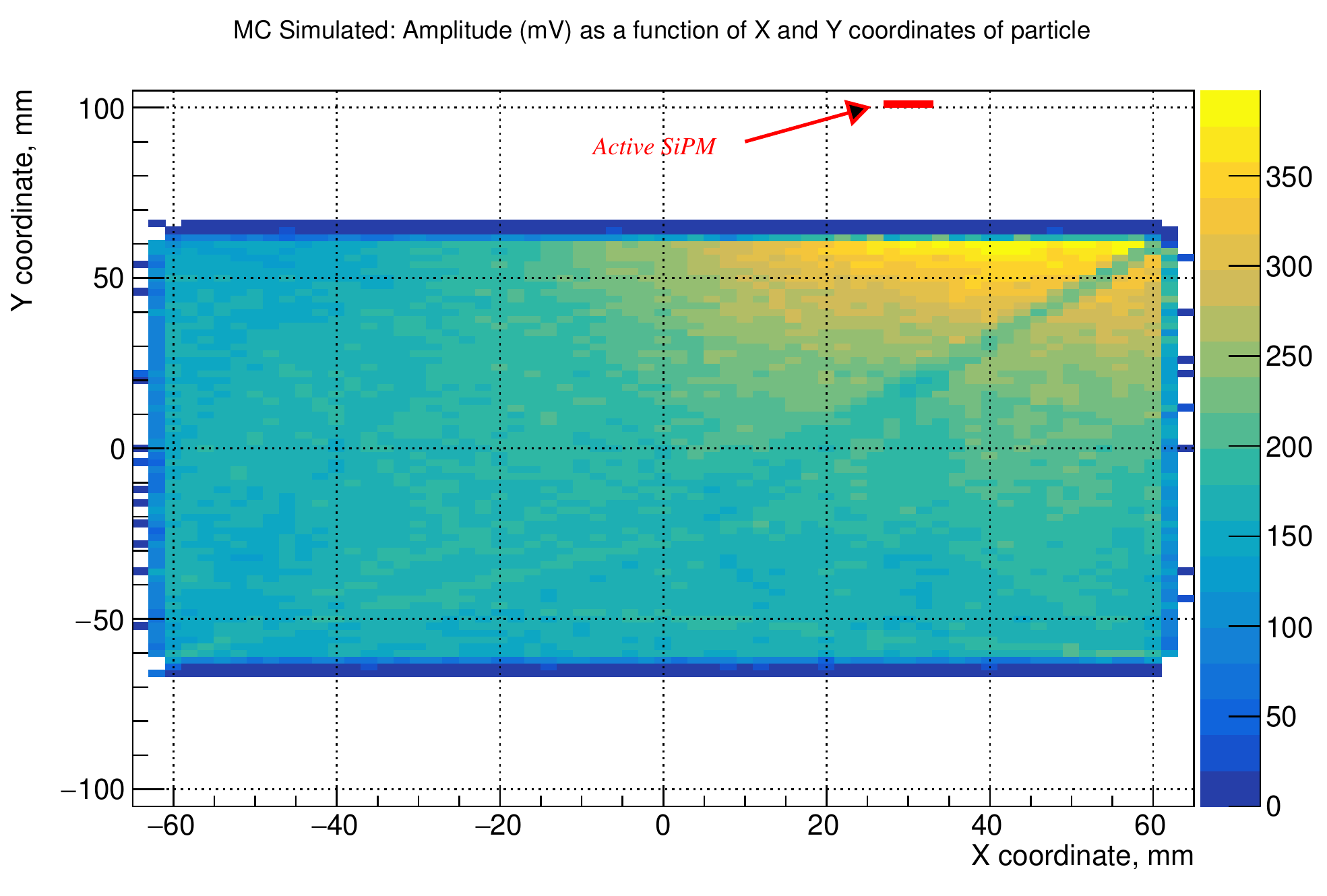}
\end{minipage}
\hfill
\begin{minipage}{0.32\textwidth}
\includegraphics[width=\linewidth]{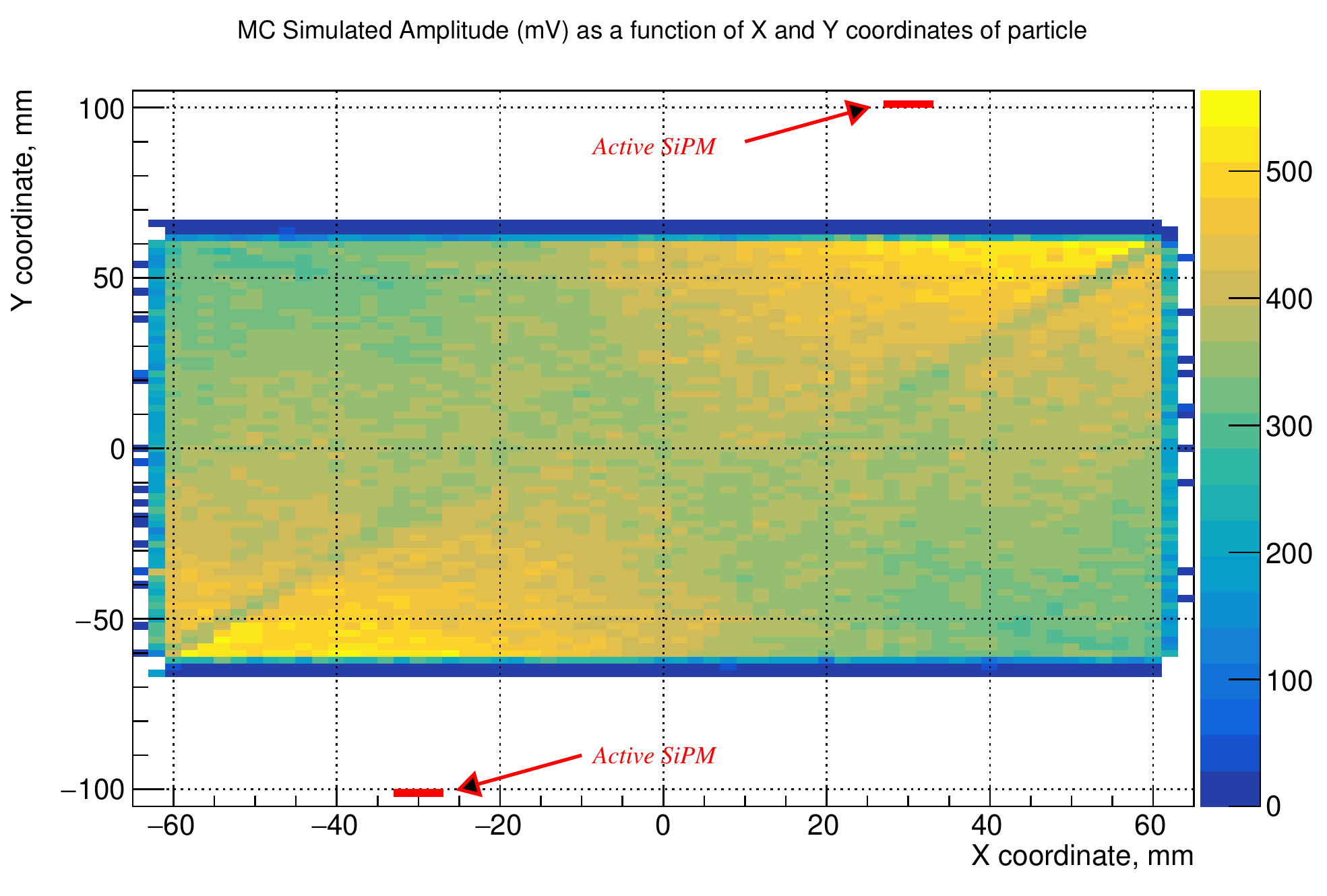}
\end{minipage}
\hfill
\begin{minipage}{0.32\textwidth}
\includegraphics[width=\linewidth]{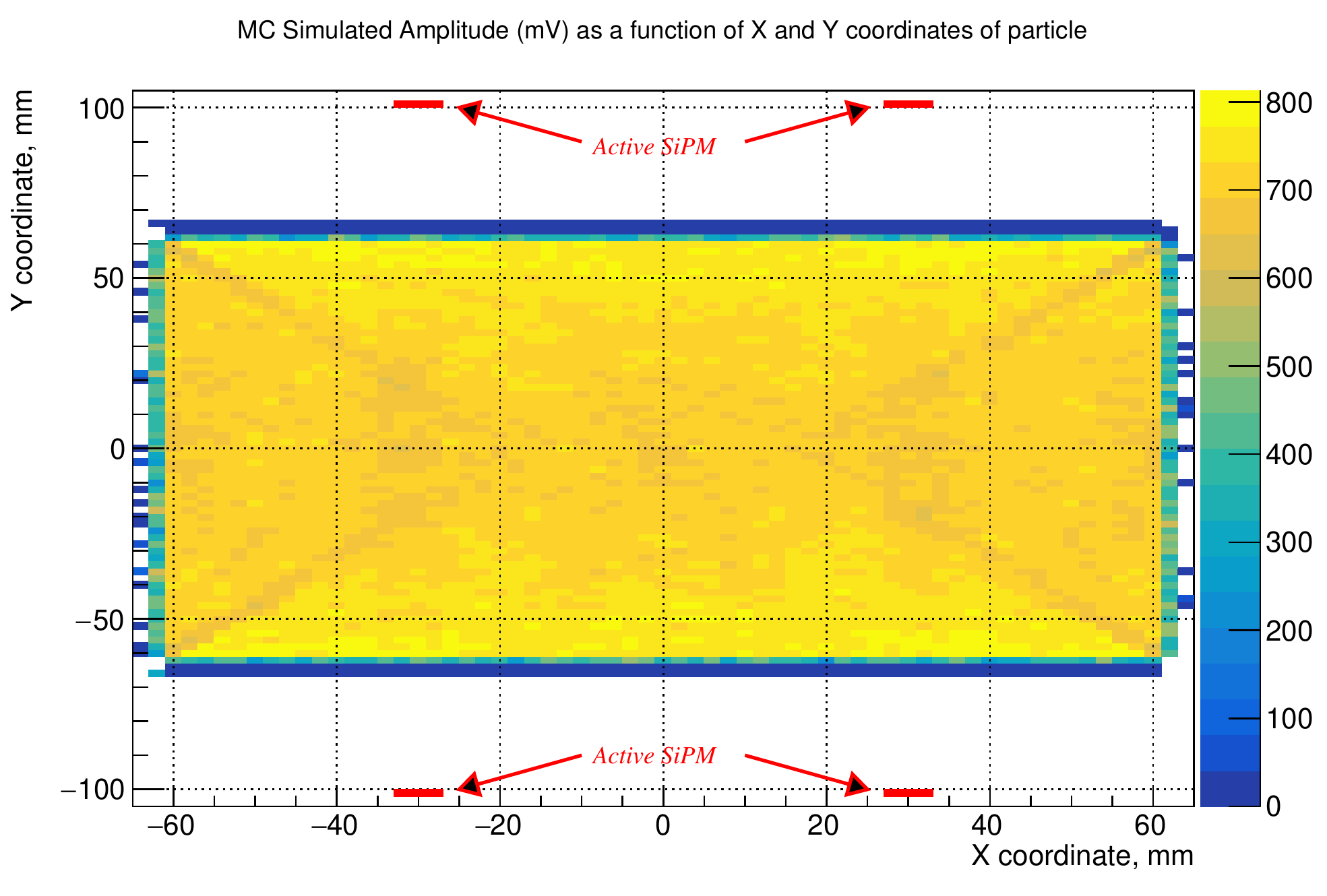}
\end{minipage}
\caption{MC simulated average pulse amplitude as a function of particle  crossing coordinates.}
\label{fig:Anti0MCAmplitude}
\end{figure}

\begin{figure}[h]
\begin{minipage}{0.32\textwidth}
\includegraphics[width=\linewidth]{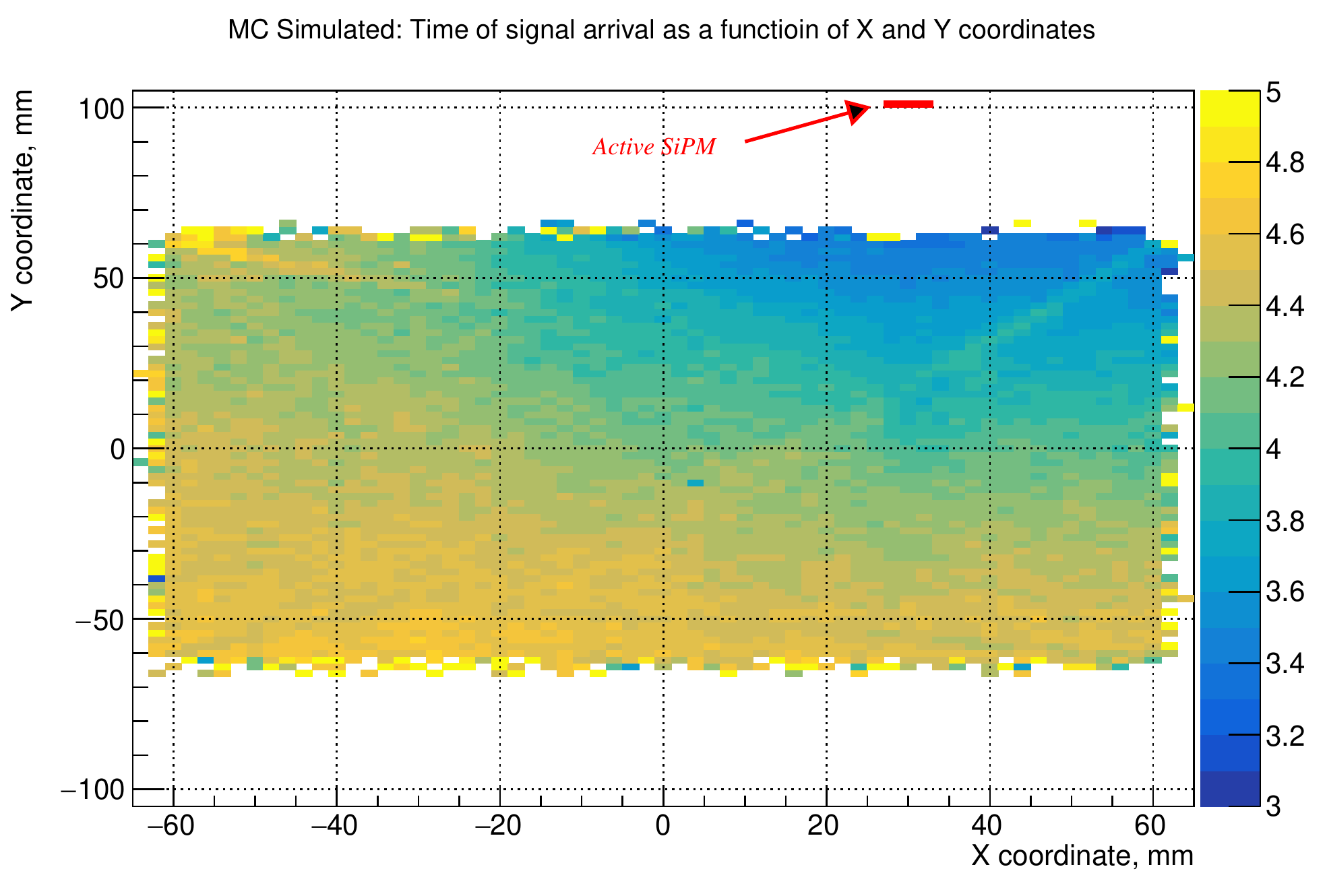}
\end{minipage}
\hfill
\begin{minipage}{0.32\textwidth}
\includegraphics[width=\linewidth]{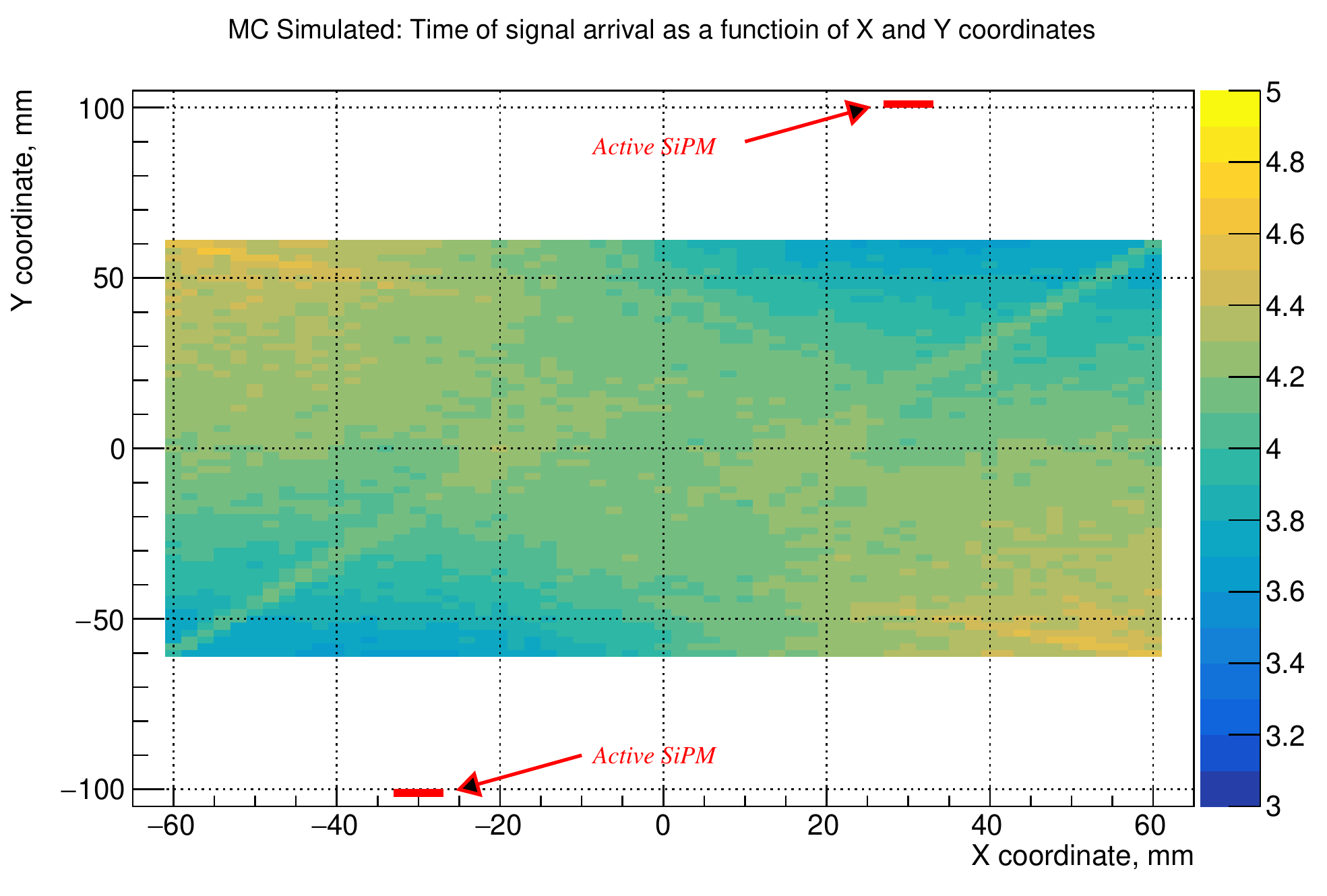}
\end{minipage}
\hfill
\begin{minipage}{0.32\textwidth}
\includegraphics[width=\linewidth]{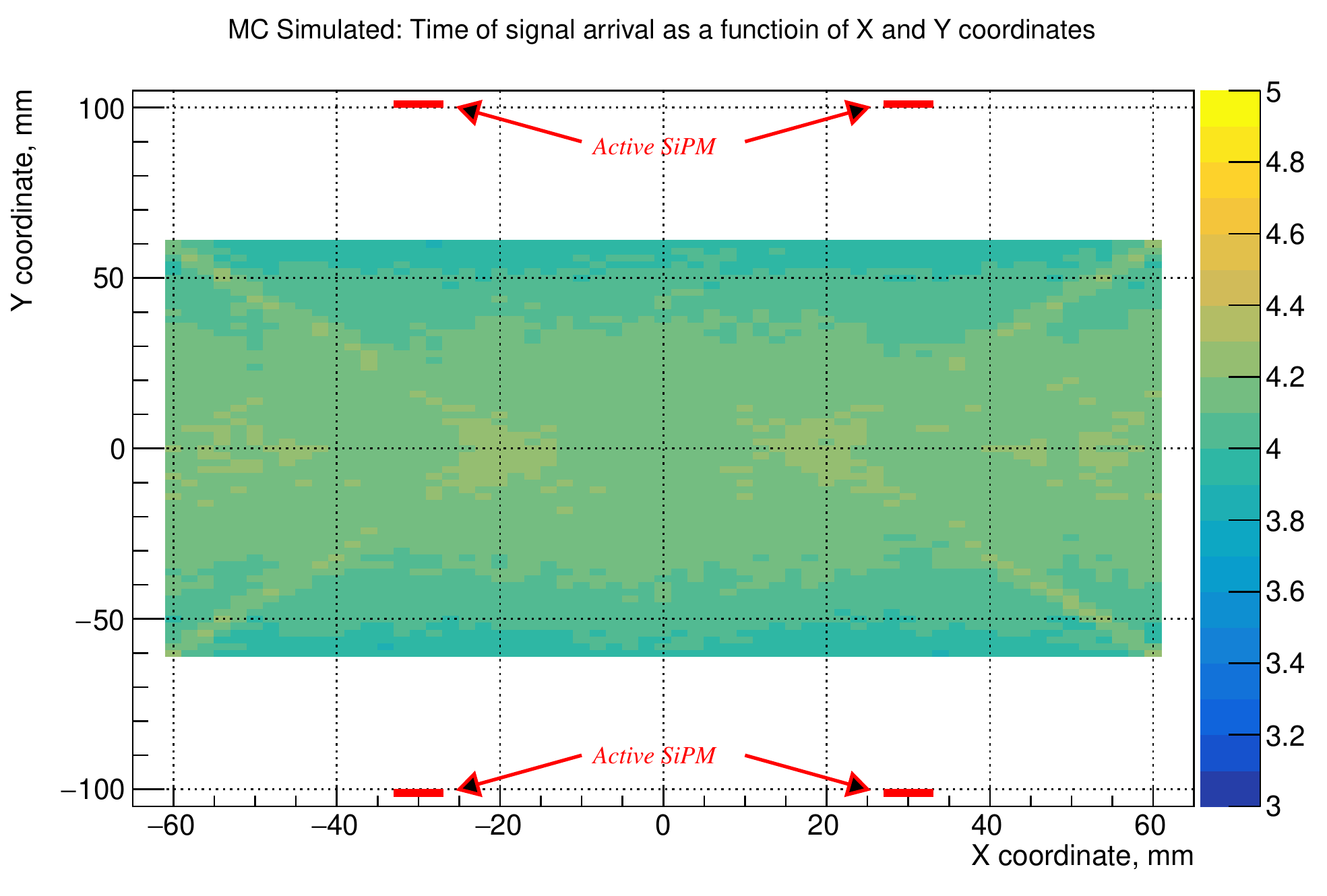}
\end{minipage}
\caption{MC simulated average time of signal arrival~(CFD with threshold = 0.2) as a function of particle crossing coordinates.}
\label{fig:Anti0MCTime}
\end{figure}

 Figures~\ref{fig:Anti0MCAmplitude} and~\ref{fig:Anti0MCTime} present the average amplitude and time of signal arrival as a function of the coordinates of the charged particle crossing the counter. The left figures correspond to the option with only one active SiPM, the middle figures correspond to the option with output pulse as a sum of the signals from two active SiPMs placed on the opposite edges of the scintillator. The right figures correspond to the output pulse being the sum of four active SiPMs.


\section{Tests with cosmic rays}
\begin{figure}[h]
\begin{minipage}{0.5\textwidth}
\centering
\includegraphics[width=0.8\linewidth]{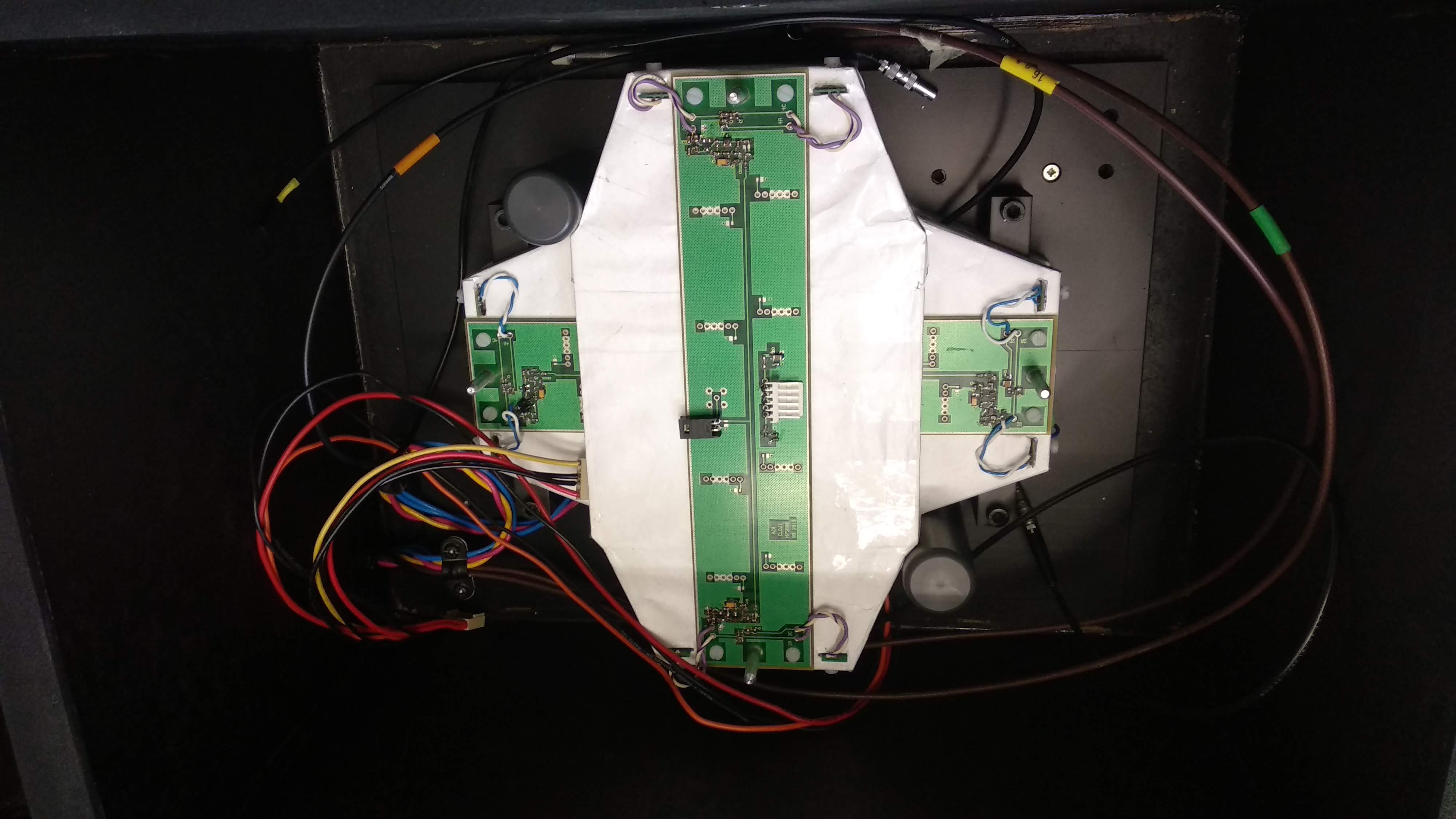}
\end{minipage}
\begin{minipage}{0.5\textwidth}
\centering
\includegraphics[width=\linewidth]{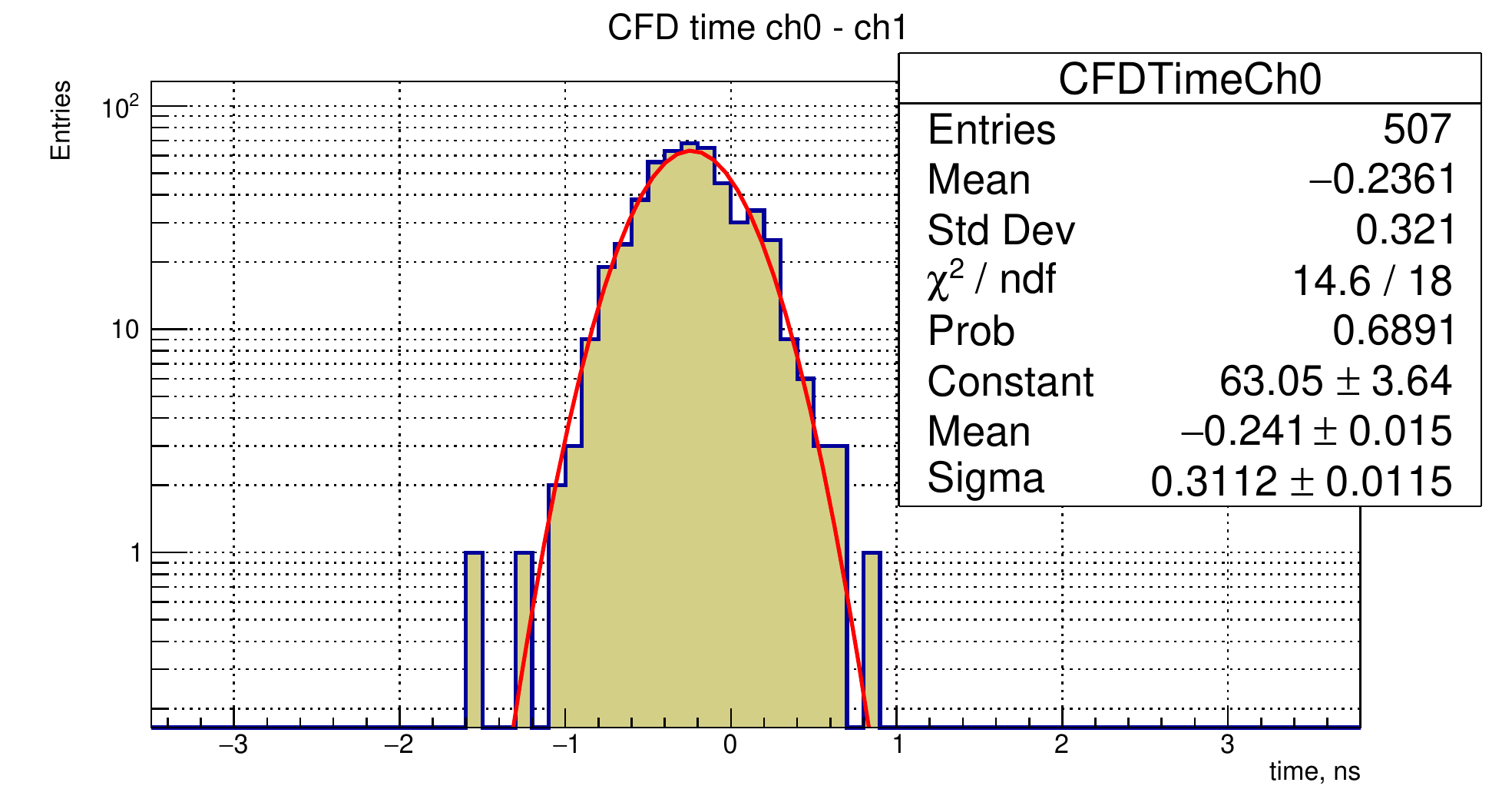}
\label{fig:Anti0Chessboard}
\end{minipage}
\caption{Test setup with the cosmic rays~(left) and the time difference between the signals arrival for the two tiles in the middle~(right) with the red line corresponding to the Gaussian fit with $\sigma = 310$~ps.}\label{fig:TestCosmics}
\end{figure}

Four tiles were placed in the light-tightened box one on top of another~(figure~\ref{fig:TestCosmics}). The coincidence of the top and bottom tile signals were used as trigger while signals from the two tiles in the middle were recorded with a digital 10~GHz oscilloscope. 
\par The difference of the signal arrival times of the central tiles~(discriminated with CFD, threshold = 0.2) is presented in figure~\ref{fig:TestCosmics}~(right). The distribution is described by a  Gaussian~(red line) with $\sigma = 310$~ps. The time resolution of the individual counter was estimated as $\sigma = 310 / \sqrt{2} = 220$~ps~(in the assumption of identical tiles). 

\section{Summary}
A new detector ANTI-0 is being assembled at CERN. The installation in the experimental hall is scheduled for the fourth quarter of 2020. Commissioning and start of the data-taking is scheduled for 2021.
 
\acknowledgments
This work is partly supported with ERC starting grant 336581.


\end{document}